\documentclass[reprint,amssymb, amsmath, aps, superscriptaddress, showpacs, footinbib, prb]{revtex4-1}
\usepackage{graphicx} 
\usepackage{amsmath}

\newcommand{\be}{\begin{equation}}
\newcommand{\ee}{\end{equation}}
\newcommand{\bea}{\begin{eqnarray}}
\newcommand{\eea}{\end{eqnarray}}
\newcommand{\bse}{\begin{subequations}}
\newcommand{\ese}{\end{subequations}}

\begin{document}

\title{Frustrated three-dimensional antiferromagnet Li$_{2}$CuW$_{2}$O$_{8}$:\\ $^7$Li NMR and the effect of non-magnetic dilution}

\author{K. M. Ranjith}
\affiliation{School of Physics, Indian Institute of Science
Education and Research Thiruvananthapuram-695016, India}
\author{M. Majumder}
\affiliation{Max Planck Institute for Chemical Physics of Solids, 01187 Dresden, Germany}
\author{M. Baenitz}
\affiliation{Max Planck Institute for Chemical Physics of Solids, 01187 Dresden, Germany}
\author{A. A. Tsirlin}
\affiliation{National Institute of Chemical Physics and Biophysics, 12618 Tallinn, Estonia}
\affiliation{Experimental Physics VI, Center for Electronic Correlations and Magnetism, Institute of Physics, University of Augsburg, 86135 Augsburg, Germany}
\author{R. Nath}
\email{rnath@iisertvm.ac.in}
\affiliation{School of Physics, Indian Institute of Science
Education and Research Thiruvananthapuram-695016, India}
\date{\today}
\begin{abstract}
We report a $^7$Li nuclear magnetic resonance (NMR) study of a frustrated three-dimensional \mbox{spin-$\frac12$} antiferromagnet Li$_{2}$CuW$_{2}$O$_{8}$ and also explore the effect of non-magnetic dilution. The magnetic long-range ordering in the parent compound at $T_{\rm N}\simeq$ 3.9~K was detected from the drastic line broadening and a peak in the spin-lattice relaxation rate ($1/T_1$). The NMR spectrum above $T_{\rm N}$ broadens systematically, and its full width at half maximum (FWHM) tracks the static spin susceptibility. From the analysis of FWHM vs. static susceptibility, the coupling between the Li nuclei and Cu$^{2+}$ ions was found to be purely dipolar in nature. The magnitude of the maximum exchange coupling constant is $J_{\rm max}/k_{\rm B}\simeq 13$\,K. NMR spectra below $T_{\rm N}$ broaden abruptly and transform into a double-horn pattern reflecting the commensurate nature of the spin structure in the ordered state. Below $T_{\rm N}$, 1/$T_1$ follows a $T^5$ behavior. The frustrated nature of the compound is confirmed by persistent magnetic correlations at high temperatures well above $T_{\rm N}$. The dilution of the spin lattice with non-magnetic Zn atoms has dramatic influence on $T_N$ that decreases exponentially similar to quasi-one-dimensional antiferromagnets, even though Li$_2$CuW$_2$O$_8$ has only a weak one-dimensional anisotropy. Heat capacity of doped samples follows power law ($C_{\rm p} \propto T^{\alpha}$) below $T_{\rm N}$, and the exponent ($\alpha$) decreases from 3 in the parent compound to 1 in the 25\% doped sample.
\end{abstract}
\maketitle

\section{Introduction}
Geometrically frustrated spin-$1/2$ triangular systems have attracted considerable attention due to numerous quantum phenomena that they exhibit at low temperatures.
The combination of geometric frustration, low spin value, and reduced dimensionality significantly enhances quantum fluctuations which can lead to the formation of exotic ground states.\cite{Balents199} The most striking example is the formation of a quantum spin liquid (QSL), which is a strongly correlated disordered ground state. Anderson suggested that the spin-$1/2$ Heisenberg triangular antiferromagnets have a resonating-valance-bond (RVB) ground state, one kind of the QSL.\cite{Anderson1973153} Indeed, several triangular antiferromagnets, such as the spin-1 compound NiGa$_2$S$_4$ and spin-$1/2$ organic materials like $k$-(BEDT-TTF)$_2$Cu$_2$(CN)$_3$ and EtMe$_3$Sb[Pd(dmit)$_2$]$_2$, do not show long-range magnetic order (LRO) and are even established as QSL candidates.\cite{Nakatsuji09092005,Yamashita04062010,Shimizu107001} On the other hand, the majority of triangular antiferromagnets entail subtle deviations from the ideal regime that may reduce the frustration and stabilize the LRO. In fact, even the idealized case of the quantum Heisenberg model on the triangular lattice features LRO with the 120$^{\circ}$ spin structure\cite{Bernu10048,Chernyshev144416} and thus lacks the spin-liquid phase. Most of the spin-$\frac12$ triangular antiferromagnets are magnetically ordered at low temperatures. They may also show plateaux at $1/3$ of the saturation magnetization, where the system evolves from a 120$^{\circ}$ magnetic ordering phase to an $up-up-down$ $(uud)$ phase in a finite field,\cite{Chubokov69} as in spin-$\frac12$ compounds Cs$_2$CuBr$_4$ and Ba$_3$CoSb$_2$O$_9$.\cite{Ono104431,Fortune257201,Susuki267201}

\begin{figure}
\includegraphics [scale = 0.7]{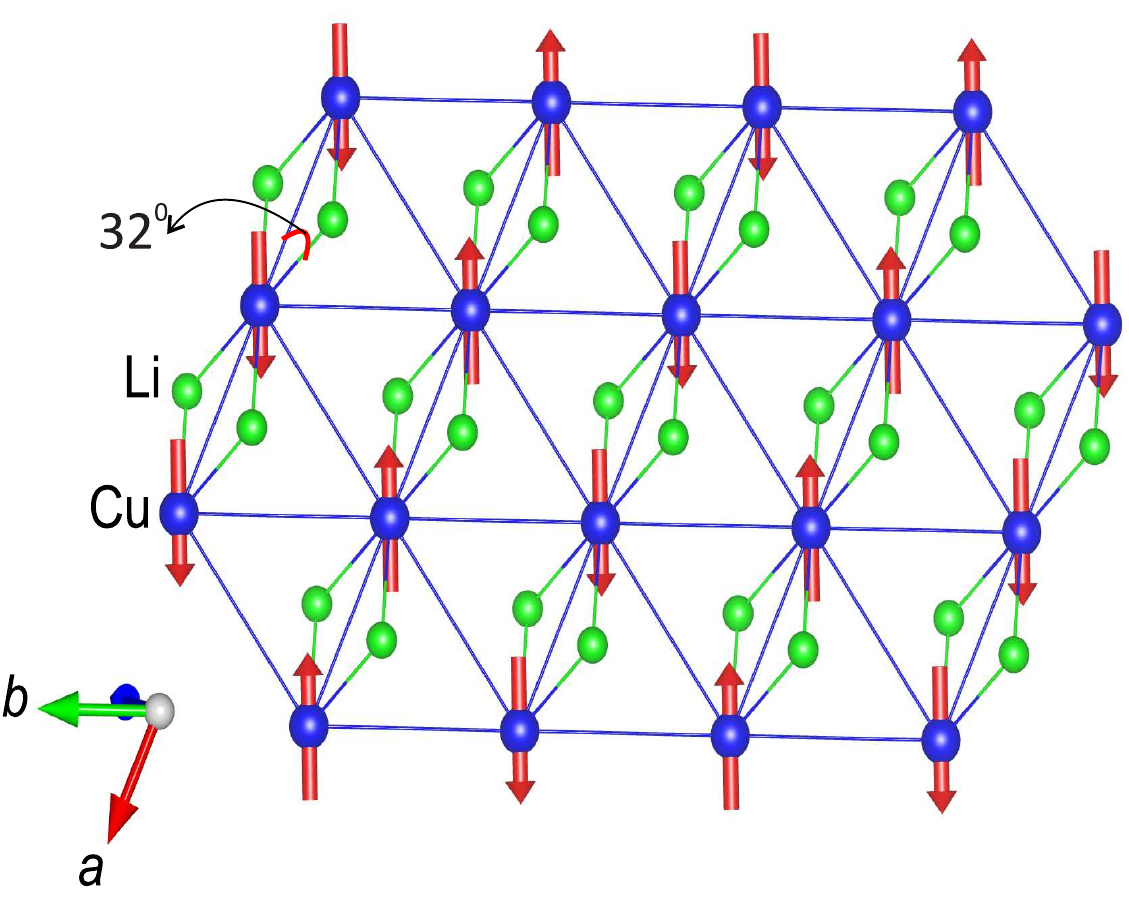}
\caption{\label{structure} The triangular layer formed by Cu$^{2+}$ ions, columnar arrangement of spins, and the couplings of Li atoms to the magnetic spins are shown.}
\end{figure}

Li$_{2}$CuW$_{2}$O$_{8}$ crystallizes in a triclinic structure with space group $P\bar{1}$ (Ref.~\onlinecite{Vega3871}) and features a unique spin lattice, which is strongly frustrated along all three crystallographic directions. The interaction topology in the $ab$ plane resembles a triangular lattice, whereas interplane couplings are oblique and likewise frustrated.\cite{Ranjith2015} There is one Cu site, two equivalent W sites, and two equivalent Li sites in the unit cell. The CuO$_6$ octahedra that can also be viewed as CuO$_4$ planar units are connected through WO$_6$ octahedra. The Li atoms are located close to each of the Cu$^{2+}$ triangular layer. Figure~\ref{structure} shows the triangular layer formed by Cu$^{2+}$ ions and the collinear arrangement of spins inferred from the neutron diffraction experiments.\cite{Ranjith2015} The Li atom is coupled to two neighbouring Cu$^{2+}$ spins (either $up-up$ or $down-down$) of the same plane.

Recently, we reported magnetic susceptibility $\chi(T)$, heat capacity $C_p(T)$, and neutron diffraction on Li$_{2}$CuW$_{2}$O$_{8}$.\cite{Ranjith2015} Clear signatures of the magnetic three-dimensionality and strong frustration were inferred from these measurements and rationalized microscopically using band-structure calculations. $\chi(T)$ and $C_p(T)$ exhibit broad maxima at around 8.5~K and 6~K, respectively, associated with the short-range order and then a kink ($\chi$) or a $\lambda$-type anomaly ($C_p$) at $T_{\rm N}\simeq$ 3.9~K associated with the magnetic long-range ordering (LRO). Neutron diffraction experiments reveal collinear magnetic order below 3.9~K with the propagation vector (0,$\frac12$,0) and the ordered moment of 0.65(4)~$\mu_{\rm B}$, which is reminiscent of a 2D antiferromagnet even though Li$_2$CuW$_2$O$_8$ is clearly three-dimensional (3D). Remarkably, the collinear order is not anticipated on the classical level, where an incommensurate spiral structure would have lower energy. Therefore, a strong stabilization of the collinear order by quantum fluctuations has been envisaged.

Li$_2$CuW$_2$O$_8$ is isostructural to its Co and Ni analogues. Both Li$_{2}$NiW$_{2}$O$_{8}$ and Li$_{2}$CoW$_{2}$O$_{8}$ compounds undergo two successive magnetic transitions at low temperatures without having any short-range order above $T_N$.\cite{Muthuselvam174430,Ranjith2015a} Many of the triangular-lattice compounds exhibit two phase transitions at low temperatures too.\cite{Markina104409,*Yokota014403,*Hwang257205,*Shirata093702} The presence of only one magnetic transition and the robust collinear order in Li$_2$CuW$_2$O$_8$ are clearly unusual and require detailed investigation.

In this paper, we report $^{7}$Li NMR measurements on Li$_{2}$CuW$_{2}$O$_{8}$ and the effect of non-magnetic Zn$^{2+}$ doping at the magnetic Cu$^{2+}$ site. Compared to the previous study where only bulk measurements were performed,\cite{Ranjith2015} we use NMR, which is a powerful local tool for studying magnetic interactions and magnetic order in frustrated spin systems. The reduction in the ordered magnetic moment (0.65\,$\mu_B$) below its classical value of 1\,$\mu_B$ has been ascribed to strong quantum fluctuations, but alternatively it could be understood as a coexistence of ordered and disordered phases below $T_N$. Our NMR study rules out this possibility and corroborates the formation of the commensurate order throughout the sample. We also demonstrate the persistence of strong spin correlations well above $T_N$, which is a hallmark of strongly frustrated magnets, and analyze spin dynamics of Li$_2$CuW$_2$O$_8$. The experiments on magnetic dilution via Zn$^{2+}$ doping bring yet another perspective on the frustrated magnetism of Li$_2$CuW$_2$O$_8$. We demonstrate an abrupt decrease in the N\'eel temperature upon doping, an effect reminiscent of quasi-one-dimensional (1D) antiferromagnets and very peculiar considering the largely 3D nature of Li$_2$CuW$_2$O$_8$.

\section{Experimental}
Polycrystalline samples of Li$_{2}$(Cu$_{1-x}$Zn$_x$)W$_{2}$O$_{8}$ ($x$ = 0, 0.05, 0.10, 0.15, 0.20, 0.25, and 0.30) were prepared by the conventional solid-state
reaction technique using Li$_{2}$CO$_{3}$ (Aldich, $99.999$\%), CuO (Aldich, $99.999$\%), ZnO (Aldich, $99.99$\%), and WO$_{3}$ (Aldich, $99.999$\%) as starting materials.
The stoichiometric mixtures were ground and fired at 650$^0$C for 24 hours to allow decarbonation and then at 700$^0$C for 48 hours with one intermediate grinding. Phase purity of the samples were checked using powder x-ray diffraction (XRD) experiment (PANalytical, Empyrean) with Cu K$\alpha$ ($\lambda$ = 1.54060\,\r A) radiation. The samples with $x$ upto 25\%, were single-phase, but at higher doping concentrations several impurity phases including Li$_2$WO$_4$ emerged. Our repeated attempts to achieve higher doping levels by increasing or lowering the firing temperature were unsuccessful. Therefore, we focus on studying the samples with $x\leq 30$\%, where a minor amount of non-magnetic impurities does not hinder the data analysis.

Temperature ($T$) dependent magnetic susceptibility $\chi(T)$ and heat capacity $C_{\rm p}(T)$ measurements were performed using a commercial Physical Property Measurement System (PPMS, Quantum Design). For the $\chi(T)$ measurement, the vibrating sample magnetometer (VSM) attachment to the PPMS was used. $C_{\rm p} (T)$ was measured by the relaxation technique on a pressed pellet using the heat capacity option of the PPMS. For the low temperature ($T<2$~K) $C_{\rm p} (T)$ measurements, a Heliun-3 attachment was used.

Nuclear magnetic resonance (NMR) experiments were carried out using pulsed NMR technique on the $^7$Li nucleus with the nuclear spin $I = 3/2$ and gyromagnetic ratio $\gamma_{\rm N}/2\pi = 16.546$~MHz/T. The NMR spectra were obtained either by Fourier transform of the NMR echo signal at a fixed field of $H = 1.5109$~T or by sweeping the magnetic field at a corresponding fixed frequency of 24.79~MHz. $^7$Li spin-lattice relaxation rate $1/T_1$ was measured using a conventional saturation pulse sequence. $^7$Li spin-spin relaxation rate $1/T_2$ was obtained by measuring the decay of the echo integral with variable spacing between the $\pi/2$ and $\pi$ pulses.

Electronic structure and electric field gradient on the Li site were calculated in the \texttt{FPLO} code\cite{Koepernik1743} using the same procedure as in Ref.~\onlinecite{Ranjith2015}.

\section{Results}
\subsection{Crystallography}
\begin{figure}
\includegraphics {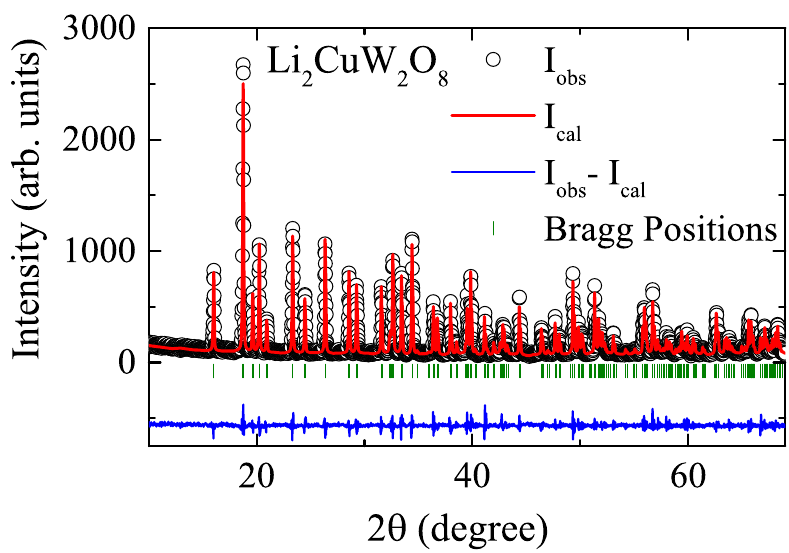}
\caption{\label{xrd} X-ray powder diffraction pattern (open
circles) at room temperature for Li$_{2}$CuW$_{2}$O$_{8}$. The solid red line represents
the calculated pattern using Le Bail fit, with the vertical bars showing the expected Bragg
peaks, and the lower solid blue line representing the difference between the
observed and calculated intensities.}
\end{figure}
Le Bail fit of the powder XRD data was performed using the \texttt{FullProf} software package based on the triclinic structure
with the space group $P\bar{1}$ to determine the lattice parameters.\cite{RodriguezCarvajal199355} All the data sets could be fitted using the structural data from Ref.~\onlinecite{Vega3871} of the parent compound as the initial parameters.
Figure~\ref{xrd} shows the room-temperature powder XRD pattern of Li$_{2}$CuW$_{2}$O$_{8}$ along with the calculated pattern. The obtained best fit parameters are listed in Table~\ref{refinement} for all the Zn doped samples. For the parent compound, the obtained lattice parameters are in close agreement with the literature data [$a=4.9669(1)$~\AA, $b=5.4969(1)$~\AA,
$c=5.8883(1)$~\AA, $\alpha=70.72(1)^{\circ}$, $\beta=85.99(1)^{\circ}$, and $\gamma=66.04(1)^{\circ}$].\cite{Vega3871}
\begin{table*}[ptb]
\centering
\caption{The refined lattice constants ($a$, $b$, and $c$), angles ($\alpha$, $\beta$, and $\gamma$), and the unit cell volume $V$ obtained from the Le Bail fit of the powder XRD data for Li$_{2}$(Cu$_{1-x}$Zn$_{x}$)W$_{2}$O$_{8}$.}
\label{refinement}
\begin{tabular}{cccccccccccc}
  \hline  \hline
$x$ & $a$ (\AA) & $b$ (\AA) & $c$ (\AA) & $\alpha(^\circ)$ & $\beta (^\circ)$ & $\gamma(^\circ)$ & $V$ (\AA$^3$)  \\\hline
0 & 4.9627(2)	& 5.4925(2)	& 5.8836(2)	& 70.717(1)	& 85.99(1)	& 66.04(1) & 	160.3729(1)\\
0.05 &	4.9596(1)	& 5.5069(1)	& 5.8892(2)	& 70.5812(9)	& 86.179(1)	& 65.917(1) & 160.8522(4)\\
0.1	& 4.9558(1)	& 5.5152(2)	& 5.8907(2)	& 70.4937(8)	& 86.3250(9)	&65.824(1) & 161.0104(2)\\
0.15	& 4.9524(2)	& 5.5218(4)	& 5.8910(2)	& 70.4180(7)	& 86.439(8)	& 65.7422(9) & 161.0986(8)\\
0.2	& 4.9487(3)	& 5.5298(4)	& 5.8918(1)	& 70.3331(8) & 86.5800(1)	& 65.6466(2) & 161.2340(8)\\
0.25	& 4.9434(2)	& 5.5353(2)	& 5.8935(1)	& 70.230(1)	& 86.723(2) & 65.55(2) & 161.2703(1)\\
0.3   & 4.9408(1) & 5.5446(1) &  5.8941 (1) &  70.147 (2) & 86.846(2) & 65.483(1) & 161.4730(1)\\
\hline
\end{tabular}
\end{table*}

In the doped samples, the lattice parameter $a$ is decreasing while $b$ and $c$ are increasing with increasing $x$ (Table~\ref{refinement}). Similarly, the angles $\alpha$ and $\gamma$ are decreasing while $\beta$ is increasing with increasing $x$. The over all unit cell volume ($V$) increases systematically, but the change in $V$ is marginal, in agreement with the fact that the ionic radius of Zn$^{2+}$ (0.6~\AA) is only slightly larger than that of Cu$^{2+}$ (0.57~\AA). 

\subsection{$^7$Li NMR spectra}
\begin{figure}
\includegraphics {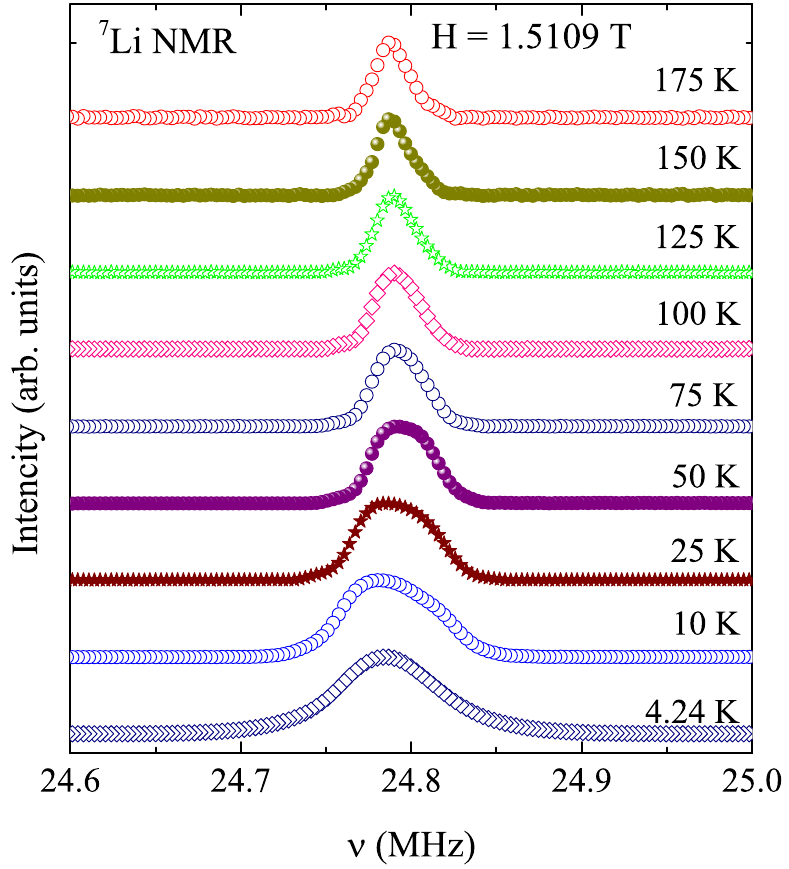}
\caption{\label{FT} $^7$Li Fourier transform spectra measured at an applied field of $H = 1.5109$~T and at different temperatures. }
\end{figure}
For the quadrupolar $^7$Li ($I = 3/2$) nucleus, one expects two satellite lines along with the central line corresponding to three allowed transitions. As shown in Fig.~\ref{FT}, we observed only a narrow single spectral line without any satellites, which is likely due to the low quadrupolar frequency or distribution of intensity of the satellites over a broad frequency range. This type of single spectral line corresponding to the Zeeman central transition ($+1/2 \longleftrightarrow 1/2$) is commonly observed in $^7$Li NMR on low-dimensional oxides.\cite{Alexander064429,Mahajan8890} Indeed, we calculated electric field gradient on the Li site and obtained a rather low value of $V_{\rm zz}=8.89\times 10^{19}$\,V/m$^2$ that yields the quadrupolar frequency of 40~kHz, which is comparable to the linewidth.

No shift in the central peak position was observed over the whole measured temperature range, above $T_{\rm N}$. The NMR shift $K$ is directly proportional to the spin susceptibility ($\chi_{\rm spin}$) and can be written as $K=\frac{A}{N_{\rm A}}\chi_{\rm spin}$, where the proportionality constant $A$ is the transferred hyperfine coupling between the $^7$Li nucleus and neighboring Cu$^{2+}$ spins, and $N_{\rm A}$ is the Avogadro's number. Thus our experimental observation of the temperature-independent line shift reflects a weak transferred hyperfine coupling of the $^{7}$Li nuclei with the Cu$^{2+}$ spins.

\begin{figure}
\includegraphics {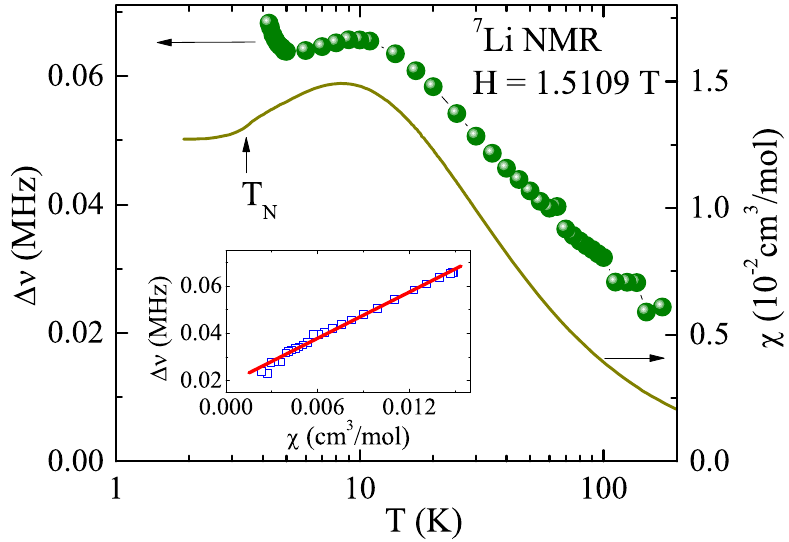}
\caption{\label{width} Temperature dependence of the full width at half maximum ($\Delta \nu$) of the $^7$Li NMR spectra measured at an applied field of $H = 1.5109$~T (left y-axis) and magnetic susceptibility ($\chi$) measured at $H = 1$~T (right y-axis). Inset shows the $\Delta \nu$ vs. $\chi$ with temperature as an implicit parameter and the solid line is the fit by Eq.~\eqref{width3}.}
\end{figure}
The spectra were found to broaden progressively upon lowering the temperature. Below about 75~K (see Fig.~\ref{FT}), the line shape becomes anisotropic. Figure~\ref{width} shows the full width at half maximum (FWHM) or $\Delta \nu$ of the $^7$Li NMR spectra as a function of temperature, above $T_{\rm N}$. $\Delta \nu$ is increasing with decreasing temperature and then shows a broad maximum at around 10~K similar to the $\chi(T)$ data.\cite{Ranjith2015} For a comparison, in the same Fig.~\ref{width}, $\chi(T)$ is also plotted in the right $y$-axis.

In a polycrystalline sample, the NMR line broadening originates from various sources. The principal sources are defects or disorder at the magnetic sites which induce staggered magnetization, anisotropy in $\chi(T)$, the dipolar interaction between the nuclei, and the dipolar hyperfine interaction of the nuclei with the neighboring magnetic ions. A small variation in bond length between the nuclei and the magnetic ions also contribute to the line broadening. In Li$_2$CuW$_2$O$_8$, since the transferred hyperfine coupling between $^7$Li nuclei and the neighboring Cu$^{2+}$ ions is negligibly small, the $^7$Li NMR line width should not be very sensitive to the staggered magnetization and the anisotropy in $\chi(T)$. Similarly, if there are distortions at the lattice site, especially in the powder sample, this leads to a small variation of inter-atomic distances. From our temperature dependent x-ray diffraction and neutron scattering experiments, no significant change in bond length between Li and Cu$^{2+}$ ions was found as a function of temperature. If at all, a small variation is there, it will not have any pronounced effect on the line width and hence ignored.

Therefore, we presumed that the shape and width of our NMR spectra are a measure of two main interactions: (i) the dipolar interaction between the nuclei and (ii) the dipolar hyperfine interaction of the $^7$Li nuclei with the neighboring Cu$^{2+}$ magnetic ions. So, one can write $\Delta \nu$ above $T_{\rm N}$ as\cite{Procissi094436,*Suh1555875,*Belesi064414}
\begin{equation}
\Delta \nu = \Delta \nu_0 + (\Delta \nu)_{\rm aniso},
\label{width1}
\end{equation}
where $\Delta \nu_0$  is the the contribution from the dipolar interaction between the nuclei, which is temperature- and field-independent. The second term $(\Delta \nu)_{\rm aniso}$ is the powder average over the anisotropic dipolar hyperfine interaction between the $^7$Li nuclei and magnetic Cu$^{2+}$ ions. In a polycrystalline sample, under magnetic field, this anisotropic coupling give rise to inhomogeneous line broadening. It is proportional to the bulk magnetic susceptibility ($\chi$) and can be written as
\begin{equation}
\frac{(\Delta \nu)_{\rm aniso}}{\nu_{\rm L}} = A_{\rm z}\chi \approx \frac{<\mu>}{r^3H},
\label{width2}
\end{equation}
where $\nu_L = \frac{\gamma_{\rm N}}{2\pi}H$ denotes the $^7$Li Larmor frequency for an applied external magnetic field $H$, $A_{\rm z}$ is the average dipolar coupling constant between the Li nucleus and magnetic Cu$^{2+}$ ions, and $r$ is the average distance between them.

It is to be noted that the shape of our experimental spectra doesn't appear to be due to what one would expect exactly from the anisotropy of the dipolar interaction. Perhaps, there are some other sources which also partly contribute to the line broadening.

Putting Eq.~\eqref{width2} in Eq.~\eqref{width1} we can write
\begin{equation}
\Delta \nu = \Delta \nu_0 + \frac{\gamma_{\rm N}}{2\pi}A_{\rm z} H\chi.
\label{width3}
\end{equation}
In the inset of Fig.~\ref{width}, $\Delta \nu$ is plotted as a function of $\chi$ with temperature as an implicit parameter. $\Delta \nu$ vs. $\chi$ follows a linear behaviour and is fitted by Eq.~\eqref{width3} in the temperature range 9~K$\leq T\leq$175~K. The parameters obtained from the fit are $\Delta \nu_0 \simeq 18.4$~KHz and $A_{\rm z} \simeq 7.83 \times$10$^{22}$\, cm$^{-3}$. This value of $A_{\rm z}$ is similar to the values reported for several other compounds\cite{Khuntia094413,*Amiri104408,*Khuntia184439} and is of the right order of magnitude with the expected dipolar interaction of Li nuclei with Cu$^{2+}$ ions at a mean distance of 2.93~\AA~apart. Thus we conclude that the hyperfine interaction of the Li nucleus with Cu$^{2+}$ moments is largely of dipolar origin.

\subsection{\textbf{$^7$Li relaxation rates ($1/T_1$ and $1/T_2$)}}
\begin{figure}
\includegraphics {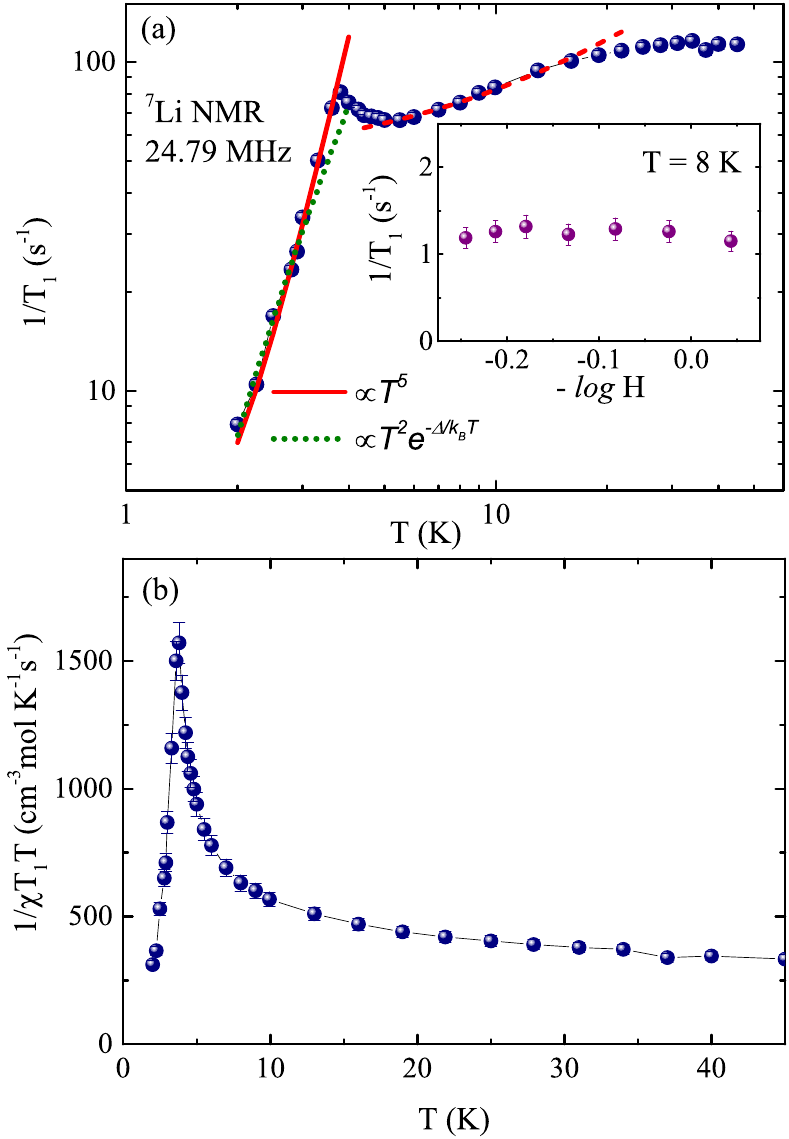}
\caption{\label{T1} (a) Temperature dependence of the $^7$Li spin-lattice relaxation rate, $1/T_1$, for Li$_2$CuW$_2$O$_8$ at $H = 1.5109$~T. The solid line and the dotted line correspond to $1/T_1 \propto T^5$ and $1/T_1 \propto T^2e^{-\Delta/k_{\rm B}T}$ behaviors, respectively. The dashed line is the linear fit ($1/T_1 \propto T$) in the $5.5~{\rm K} \leq T \leq 16$~K range. The inset shows $1/T_1$ vs. $-\log H$ measured at $T = 8$~K. (b) $1/\chi T_1T$ as a function of temperature.}
\end{figure}
The recovery of the longitudinal magnetization after a saturation pulse was fitted well by a single exponential function
\begin{equation}
1-\frac{M(t)}{M_0} = A e^{-t/T_1},
\label{expo}
\end{equation}
where $M(t)$ is the nuclear magnetization at a time $t$ after the saturation pulse and $M_0$ is the equilibrium value of the magnetization. The temperature dependence of $1/T_1$ extracted from the fitting is shown in Fig~\ref{T1}(a). At high temperatures ($T\geq 20~K$), 1/$T_1$ is almost temperature-independent (in the paramagnetic regime), which is often observed in a system with localized moments when the temperature is higher than the exchange energy between the spins.\cite{Moriya01071956} With decreasing temperatures, 1/$T_1$ decreases almost linearly for $T < 20~K$ where the compound starts approaching the short-range-ordered regime ($T\simeq$ 10~K). With further decrease in temperature, a weak but discernible anomaly was observed at $T_N\simeq 3.6$~K due to the critical slowing down of fluctuating moments while approaching the long-range magnetic ordering, nearly consistent with the thermodynamic\cite{Ranjith2015} measurements ($T_{\rm N} \simeq 3.9$~K).\footnote{The small discrepancy in $T_{\rm N}$ between NMR and thermodynamic measurements is due to the different thermocouples used in two different experiments.} Below $T_{\rm N}$, $1/T_1$ decreases rapidly towards zero as a result of the disappearance of the critical fluctuations in the ordered state.


We also measured 1/$T_1$ at $T = 8$~K for different applied fields in order to check the effect of spin diffusion. It is known that in Heisenberg magnets, spin correlation function exhibits a diffusive effect due to the contribution of long wavelength or small wave vector ($q \sim 0$).\cite{Van1374} In one-dimensional (1D) spin systems, spin diffusion shows a field dependence of 1/$T_1 \propto$ 1/$\sqrt{H}$ type. Similar type of field dependence is reported in Heisenberg spin-chain compounds (CH$_3$)$_4$NMnCl$_3$, AgVP$_2$S$_6$, CuCl$_2\cdot$2NC$_5$H$_5$, and Sr$_2$CuO$_3$.\cite{Borsa2215,Hone965,Takigawa2173,Ajiro420,Thurber247202} On the other hand, in two-dimensional (2D) systems, spin diffusion results in a $\log$(1/$H$) dependence of 1/$T_1$, which has been experimentally observed in the quasi-2D Heisenberg magnet Zn$_2$VO(PO$_4$)$_2$.\cite{Furukawa2393,Yogi024413} As presented in the inset of Fig.~\ref{T1}(a), $1/T_1$ for Li$_2$CuW$_2$O$_8$ at $T = 8$~K is almost independent of the applied field confirming the absence of diffusive dynamics.

\begin{figure}
\includegraphics {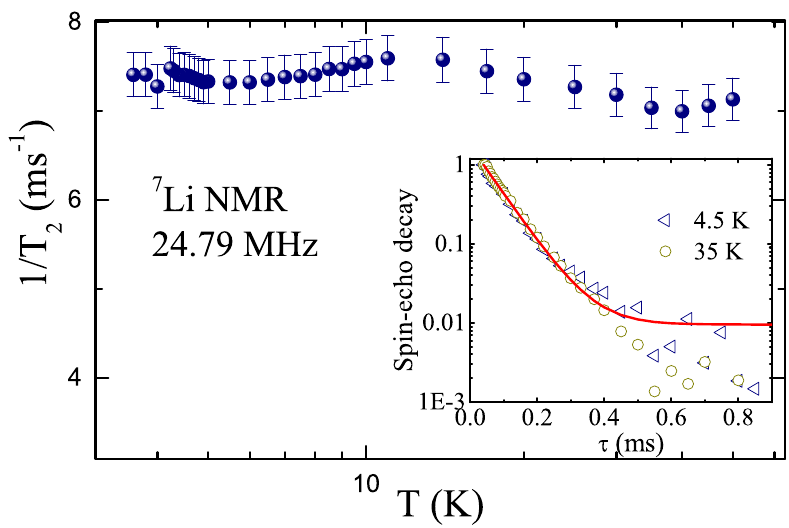}
\caption{\label{T2} $1/T_2$ as a function of temperature. Inset: Spin-echo decays are plotted as a function of $\tau$ at two different temperatures. The solid line is the fit using Eq.~\eqref{expo}.}
\end{figure}
The spin-spin relaxation rate $1/T_2$ was obtained by monitoring the decay of the transverse magnetization ($M_{\rm xy}$) after a $\pi/2 - \tau - \pi$ pulse sequence as a function of the pulse separation time $\tau$ and fitting by the following equation
\begin{equation}
M_{\rm xy} = M_0 e^{-2\tau/T_2} + C.
\label{expo}
\end{equation}
The extracted $1/T_2$ is plotted as a function of temperature in Fig.~\ref{T2} and is found to be almost temperature independent.

\subsection{\textbf{$^7$Li NMR spectra below $T_{\rm N}$}}
\begin{figure}
\includegraphics {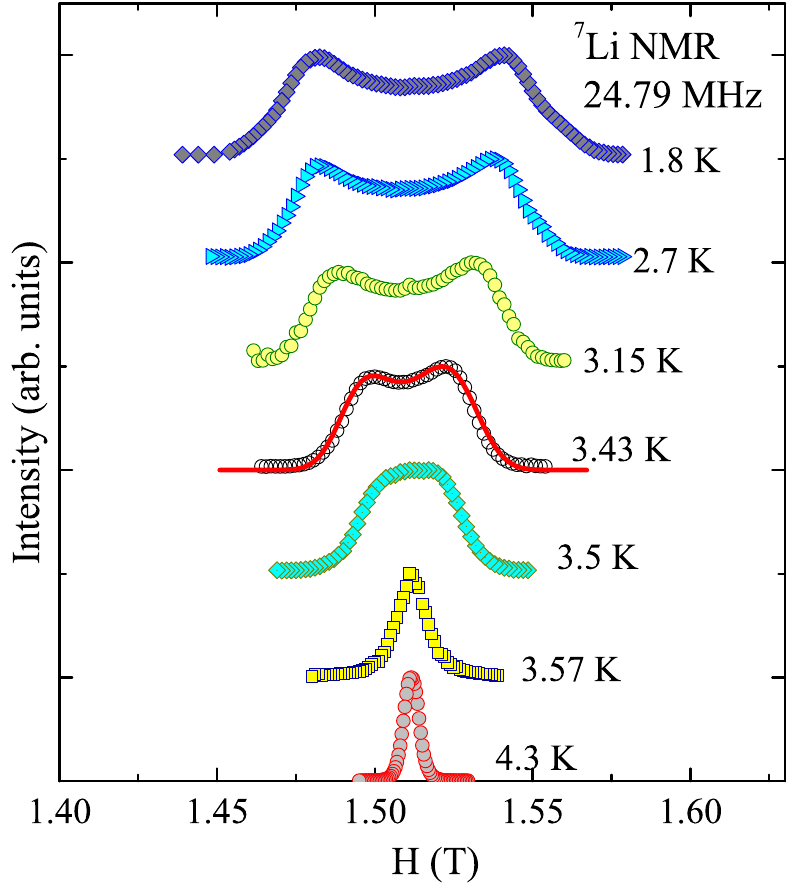}
\caption{\label{sweep} Temperature-dependent field-sweep $^7$Li NMR spectra measured at 24.79~MHz below $T_{\rm N}$. The red solid line is the fit to the spectra at 3.43~K using the sum of two Gaussian functions.}
\end{figure}
Field-sweep $^7$Li NMR spectra of Li$_2$CuW$_2$O$_8$ below $T_{\rm N}$ are shown in Fig.~\ref{sweep}. As discussed before, the NMR line above $T_{\rm N}$ broadens systematically with decreasing temperature. While approaching $T_{\rm N}$, it broadens abruptly and then transforms into a nearly rectangular pattern well below $T_{\rm N}$. The broadening of the spectra below $T_{\rm N}$ indicates that the Li site is experiencing the static internal field in the ordered state through the hyperfine coupling between the $^7$Li nuclei and the ordered Cu$^{2+}$ moments. At low temperatures, the rectangular spectra develop two horns on either side of its edges. The spectral width is found to be increasing progressively upon lowering the temperature.

\begin{figure}
\includegraphics {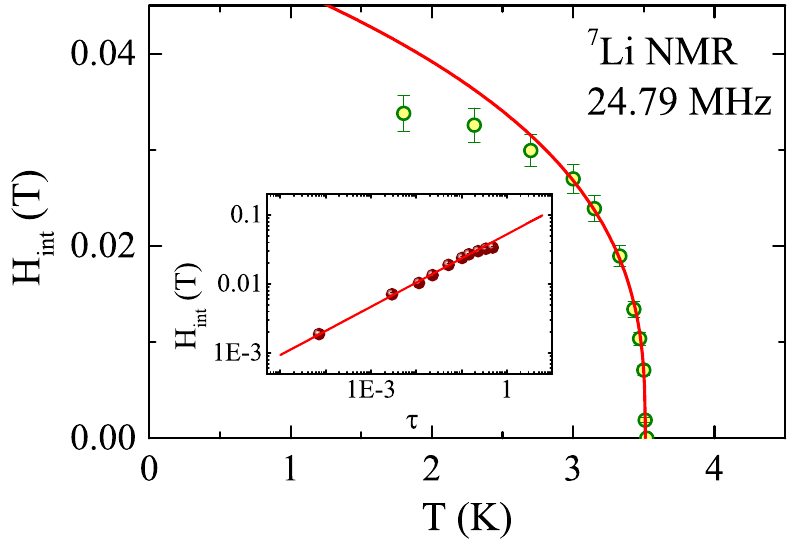}
\caption{\label{Hn} Temperature dependence of the internal field $H_{\rm int}$ obtained from $^7$Li NMR spectra
measured at 24.79~MHz in the ordered state. The solid line is the fit by Eq.~\eqref{Hint} as described in the text.
Inset: $H_{\rm int}$ vs. $\tau$ and the solid line is the simulation of $0.053 \times \tau^{0.35}$ taking $T_{\rm N} \simeq 3.51$~K.}
\end{figure}
The internal field $H_{\rm int}$ below $T_{\rm N}$, which is proportional to the Cu$^{2+}$ sublattice magnetization, was determined by taking half of the FWHM of the spectral lines. The temperature dependence of $H_{\rm int}$ after subtracting the high temperature ($T>T_{\rm N}$) paramagnetic line width is shown in Fig.~\ref{Hn}. In order to extract the critical exponent ($\beta$) of the order parameter (sublattice magnetization), $H_{\rm int} (T)$ was fitted by the power law
\begin{equation}
H_{\rm int}(T) = H_{\rm 0}\left(1-\frac{T}{T_{\rm N}}\right)^\beta.
\label{Hint}
\end{equation}
It is noticed that $H_{\rm int}$ decreases sharply on approaching $T_{\rm N}$. To get a reliable estimation of the critical exponent $\beta$, one needs more data points close to $T_{\rm N}$. Therefore, we have measured spectra in small temperature steps just below $T_{\rm N}$ (i.e. in the critical region). Our fit by Eq.~\eqref{Hint} in the temperature range from 3~K to 3.6~K (see Fig.~\ref{Hn}) yields $\beta$ = 0.35(2) with $H_0 = 0.053(2)$~T and $T_{\rm N}$ = 3.51(1)~K. In order to magnify the fit in the critical region, $H_{\rm int}$ is plotted against the reduced temperature $\tau = \frac{\vert T-T_{\rm N}\vert}{T_{\rm N}}$ in the inset of Fig.~\ref{Hn}. The solid line is the fit by $0.053 \times \tau^{0.35}$ where $T_{\rm N}$ is fixed to the value 3.51~K. At low temperatures, the $H_{\rm int}$ value almost reaches saturation but much faster than the mean-field prediction (see the deviation between hollow spheres and solid line in Fig.~\ref{Hn}). This behavior resembles temperature evolution of the ordered magnetic moment determined by neutron diffraction.\cite{Ranjith2015} However, the neutron data are less accurate in this case because of difficulties in measuring very weak magnetic reflections in the vicinity of $T_{\rm N}$, especially in a powder experiment.

The value of $\beta$ obtained from the analysis of sub-lattice magnetization reflects the nature of the critical fluctuations and hence the universality class or spin dimensionality of the spin systems. For a purely 3D ordering, the correlations in the critical
regime, very close to $T_{\rm N}$, will reflect 3D character. A
tiny in-plane anisotropy can dominate the fluctuations if the
dominant AF correlation length is large. If such a 3D ordering is still governed by the 2D
processes, the phase boundaries that are observed should reflect the underlying 2D character.
For instance, in BaNi$_2$(PO$_4$)$_2$, $\beta \simeq 0.33$ close to 3D model was found near $T_{\rm N}$ and another exponent $\beta \simeq 0.23$ close
to that of the 2D XY model was found in a broader region below $T_{\rm N}$.\cite{Benner}
However, in some cases this 3D critical region is over
such a narrow temperature range that it is not accessible
experimentally. Such a scenario has been realized before in
the frustrated AFM square-lattice compounds
Pb$_2$VO(PO$_4$)$_2$ and Li$_2$VOSiO$_4$. In these compounds, neutron diffraction
experiments show that the ordering is 3D. In contrast, a reduced value of $\beta$ close to the value
of the 2D XY model has been reported from NMR measurements
and the authors suggested that the transition to the
columnar phase might be driven by the XY type anisotropy.\cite{Nath214430,Melzi024409,*Melzi1318} The $\beta$ values expected for different universality classes (spin and spatial dimensionalities) are tabulated in Ref.~\onlinecite{Nath214430}. Our obtained value of $\beta \simeq 0.35(2)$ is close to the value expected for any of the 3D spin models (Heisenberg, Ising or XY). This also reflects dominant 3D correlations in the critical
regime and is consistent with our neutron diffraction experiments that predict the 3D ordering below $T_{\rm N}$.


\subsection{Zn doping}
\begin{figure}
\includegraphics{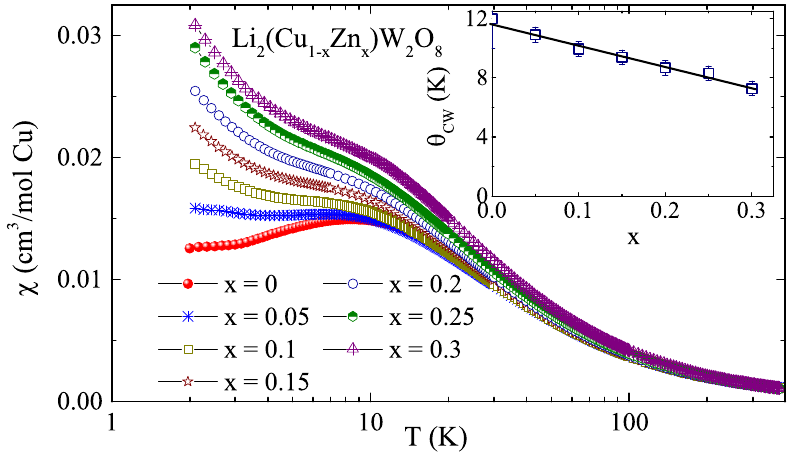}
\caption{\label{chi-doped} (Color online) $\chi(T)$ of Zn-doped Li$_2$CuW$_2$O$_8$ samples measured at $\mu_0H=1$\,T. Inset: Curie-Weiss temperature $\theta_{\rm CW}$ as a function of doping concentration $x$ and the solid line is a linear fit.}
\end{figure}

As mentioned in Sec.~II, all Zn-doped samples crystallize in triclinic symmetry, similar to the parent compound. The sample with $x=0.30$ contained trace amounts of impurity phases, so its doping level may be slightly below 30\%, but this minor deviation did not have any visible effect on the results.

$\chi(T)$ of doped samples normalized to one mole of Cu$^{2+}$ spins is shown in Fig.~\ref{chi-doped}. The susceptibility maximum is systematically shifted to higher values of $\chi$ and almost temperature-independent, although it is gradually smeared out by the low-temperature susceptibility upturn that becomes more pronounced at higher doping levels because of the increasing amount of paramagnetic spins. The weak kink at $T_{\rm N}$ vanishes upon doping. However, the magnetic transition is still clearly seen in the heat-capaciy data presented below. 

The high-temperature data, $\chi(T)$, was fitted by
\begin{equation}
\chi(T) = \chi_{0}+\frac{C}{T+\theta_{\rm CW}},
\label{CW}
\end{equation}
where $\chi_{0}$ is the temperature-independent susceptibility and the second term is the Curie-Weiss (CW) law with $C$ being the CW constant and $\theta_{\rm CW}$ the CW temperature. From the fit above 50~K, the value of $C$ was found to vary from 0.362~cm$^3$~K/mol~Cu to 0.454~cm$^3$~K/mol~Cu which correspond to the effective moment of $\mu_{\rm eff} = 1.7 - 1.9~\mu_{\rm B}$. The extracted values of $\theta_{\rm CW}$ decrease linearly upon doping (see the inset of Fig.~\ref{chi-doped}) from 12~K for the parent compound to 7.3~K for the $x=0.30$ sample. This reduction in $\theta_{\rm CW}$ confirms the dilution of the spin lattice. For random dilution, one expects that the slope of $\theta_{\rm CW}(x)/\theta_{\rm CW}(x=0)$ vs. $T$ is 1.0, which is comparable to the slope of 1.3 extracted from our data.

\begin{figure}
\includegraphics{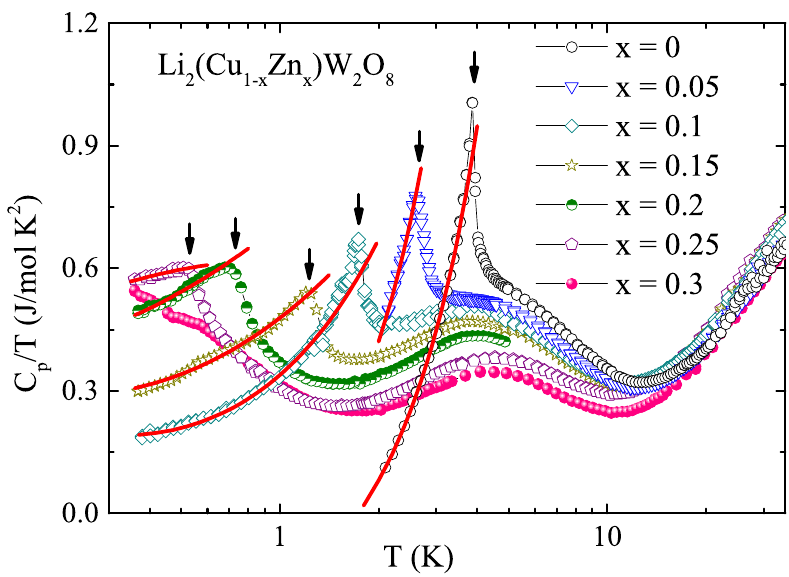}
\caption{\label{heat-doped} (Color online) $C_{p}/T$ of Zn-doped Li$_2$CuW$_{2}$O$_8$ samples measured in zero magnetic field. The downward arrows point to the N\'eel temperatures $T_N$. The solid lines are the fits by power law below $T_{\rm N}$.}
\end{figure}

For the sake of better presentation, we use a different scaling for the heat capacity and normalize the data to one mole of the compound. Figure~\ref{heat-doped} presents $C_{\rm p}/T$ vs. $T$ for all the doped samples. The systematic reduction in the heat capacity maximum around 6~K follows the reduced amount of the magnetic Cu$^{2+}$ ions. The position of the broad specific-heat maximum, which is only partially visible at $x=0$ and becomes very pronounced at $x=0.3$, is roughly unchanged up to the highest possible doping level.

Magnetic ordering transition is clearly seen in the heat-capacity data. The N\'eel temperature determined with the 0.05\,K uncertainty from the maximum of the transition anomaly, displays a systematic reduction from 3.9\,K in the parent compound to 530~mK at $x=0.25$ and eventually below 350~mK at $x=0.30$. Below $T_{\rm N}$, $C_{\rm p}(T)$ data were fitted by the power law $C_{\rm p} \propto T^{\alpha}$ for all the doped samples, where $\alpha$ is the exponent. The $\alpha$ parameter is also strongly dependent on the doping level and decreases from $\alpha=3$ at $x=0$ to $\alpha\simeq 1$ at $x=0.25$ (see Fig.~\ref{Tn}).

\section{Discussion}
\subsection{Static properties}
The broadening of the spectra below $T_{\rm N}$ suggests that a net field exists at the Li-site due to the nearest neighbor Cu$^{2+}$ spins.
As demonstrated in Fig.~\ref{structure}, Li atoms are located slightly off the plane (at an angle $\sim 32^0$) but equidistant from two neighboring spins and each Li is coupled to two neighboring spins having the same direction (either both up or both down). In such a scenario, one expects a strong internal field since the hyperfine fields due to the two up- or down- spins will be added up at the Li site. This is possible only if Li is coupled strongly to the Cu$^{2+}$ ions. As shown in Fig.~\ref{Hn}, the internal field at 1.8~K is about 0.038~T, which is relatively weak given the fact that Li is coupled to two spins having the same direction and, thus producing same direction of the local field at the Li site. This weak internal field is comparable to the value reported in $^{31}$P NMR for Zn$_2$VO(PO$_4$)$_2$, where P is strongly coupled to one up- and one down- spins, so that the resulting internal field is only due to the difference in the hyperfine couplings along different $^{31}$P--V$^{4+}$ pathways.\cite{Yogi024413} Such a comparison proves that even though Li is coupled to two spins having the same direction, the diminutively small transferred hyperfine coupling results in only a weak static field. The weak hyperfine coupling is also confirmed by the temperature-independent line shift at high temperatures ($T>T_{\rm N}$). Our spectral width analysis indeed suggests that the coupling between Li and Cu atoms is dipolar in nature, and the overlap of the $s$-orbital wave function of Li with the $3d$ wave function of Cu$^{2+}$ is almost negligible.

For a powder sample in the magnetically ordered state, the direction of the internal field $H_{\rm int}$ is distributed randomly in all directions. When the external field ($H$) is applied, the effective field $H_{\rm eff}$ at a nuclear site is $H_{\rm int}+H$. In a ferromagnetic sample, $H_{\rm eff}$ is usually parallel to $H$ and has a unique value, $H_{\rm int}+H$ or $\vert H_{\rm int}-H \vert$ in its magnitude, if $H_{\rm int}$ is unique. This is because the ferromagnetic moments are rotated easily to the direction of $H$ in a relatively weak field. On the other hand, in case of an antiferromagnet, the moments hardly change their directions in the external field in usual NMR experiments. The direction of $H_{\rm int}$ has the random distribution even under external fields. Thus, $H_{\rm eff}$ has a distribution in its direction and its magnitude varies from $\vert H_{\rm int}-H \vert$ to $H_{\rm int}+H$.
In such a case one expects a typical rectangular powder spectra in the commensurate AFM ordered state which has been experimentally observed in compounds like CuV$_2$O$_6$, BaCo$_2$V$_2$O$_8$, and BiMn$_2$PO$_6$.\cite{Kikuchi2660,Yamada1751,Ideta094433,Nath024431}

Our $^{7}$Li NMR spectra however exhibit two peaks on the top of a rectangular background. Such a spectrum with two horns has often been observed in NMR experiments on powder samples. For instance, powder NMR spectra of YMn$_2$ show a double horn pattern in the ordered state where Mn moments are coupled antiferromagnetically and collinearly.\cite{Yoshimura198355} Recently, the powder spectra of Zn$_2$VO(PO$_4$)$_2$ in the commensurate AFM ordered state showed similar pattern with two horns on top of a rectangular background.\cite{Yogi024413} According to Yamada \textit{et al.},\cite{Yamada1751} the shape of the spectra is also sensitive to instrumental features. When the axis of the sample coil of the NMR probe head, which is usually
set perpendicular to the external field $H$, is not perpendicular to the distributed directions of $H_{\rm eff}$'s then the intensity of NMR signal from the nuclei under investigation is reduced. In this case, the spectrum is curved around the middle having two peaks on either sides. Thus the double horn pattern in Fig.~\ref{sweep} is reminiscent of a commensurate AFM order and is consistent with the neutron diffraction results.\cite{Ranjith2015}

As found from the neutron diffraction experiments,\cite{Ranjith2015} the ordered magnetic moment below $T_N$ is reduced to 0.65\,$\mu_B$ from its classical value of 1\,$\mu_B$ which has been attributed to strong quantum fluctuations. Alternatively, it can be interpreted as a coexistence of ordered and disordered phases below $T_N$. In such a case, a narrow central line corresponding to disordered spins would have been observed at the zero-shift position on top of the rectangular back ground. The ratio of area under the narrow line with the area under the broad back ground would give information about the fraction of the disordered phase below $T_{\rm N}$. Our experimental spectra below $T_{\rm N}$ don't show any narrow central peak suggesting that the ordered phase is purely collinear below $T_{\rm N}$, without any signature of disorder.

\subsection{Dynamic properties}
Generally, one can express $\frac{1}{T_1T}$ in terms of dynamic susceptibility $\chi(\vec{q},\omega)$ as\cite{Moriya01071956,Mahajan8890}
\begin{equation}
\frac{1}{T_{1}T} = \frac{2\gamma_{N}^{2}k_{B}}{N_{\rm A}^{2}}
\sum\limits_{\vec{q}}\mid A(\vec{q})\mid
^{2}\frac{\chi^{''}(\vec{q},\omega_{0})}{\omega_{0}},
\label{t1form}
\end{equation}
where the sum is over the wave vectors $\vec{q}$ within the first Brillouin zone,
$A(\vec{q})$ is the form factor of the hyperfine interactions as a
function of $\vec{q}$, and
$\chi^{''}_{M}(\vec{q},\omega _{0})$ is the imaginary part of the
dynamic susceptibility at the nuclear Larmor frequency $\omega _{0}$.
For $q=0$ and $\omega_{0}=0$, the real component of $\chi_{M}^{'}(\vec{q},\omega _{0})$ corresponds to the uniform static susceptibility $\chi$. Therefore, one would expect 1/$\chi T_1T$ to be temperature-independent in the paramagnetic regime.
As shown in Fig.~\ref{T1}(b), 1/$\chi T_1T$ is almost temperature-independent above 30~K as expected. For $T\leq$ 30~K, it increases slowly and shows a sharp increase below 10~K. This increase in 1/$\chi T_1T$ implies the dominance of $\chi^{''}_{M}(\vec{q},\omega _{0})$ over $\chi(0,0)$, which is due to the growth of spin fluctuations with $q\neq 0$, even at temperatures much higher than $T_{\rm N}$. So, we can infer that strong antiferromagnetic spin fluctuations persist in a wide temperature range above $T_{\rm N}$. This indicates the frustrated nature of the Li$_2$CuW$_2$O$_8$ in agreement with thermodynamic measurements.\cite{Ranjith2015}
\begin{figure}
\includegraphics {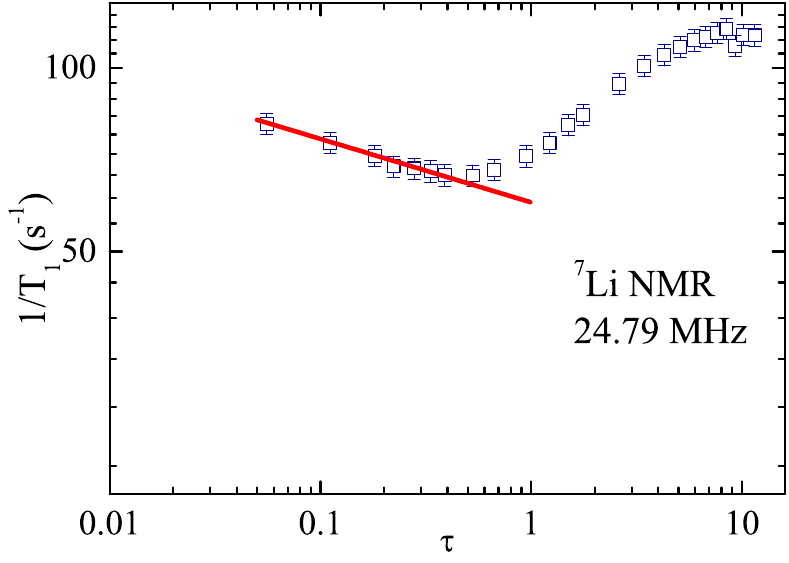}
\caption{\label{T-Tn} $^7$Li $1/T_1$ for Li$_2$CuW$_2$O$_8$ vs. the reduced temperature ($\tau$). The solid line is a fit by the power law, 1/$T_1 \propto \tau^{-\gamma}$ with $\gamma\simeq0.1$ and a fixed $T_{\rm N} \simeq 3.6$~K.}
\end{figure}

Since the dipolar coupling between Li and Cu atoms is known, it is possible to calculate the leading exchange coupling between Cu$^{2+}$ ions from the high-temperature value of $1/T_1$. At high temperatures, $1/T_1$ is constant and can be expressed as:\cite{Moriya01071956}
\begin{equation}
\left(\frac{1}{T_1}\right)_{T\rightarrow\infty} =
\frac{(\gamma_{N} g\mu_{\rm B})^{2}\sqrt{2\pi}z^\prime S(S+1)}{3\,\omega_{ex}}
{\Big(\frac{A_{z}}{z'}\Big)^{2}},
\label{t1inf}
\end{equation}
where $\omega_{ex}=\left(|J_{\rm max}|k_{\rm B}/\hbar\right)\sqrt{2zS(S+1)/3}$ is the Heisenberg exchange frequency, $z$ is the number of nearest-neighbor spins of each Cu$^{2+}$ ion, and $z^\prime$ is the number of nearest-neighbor Cu$^{2+}$ spins for a given Li site. The value of $A_{\rm z} \simeq 7.83 \times 10^{22}$~cm$^{-3}$ corresponds to 1.1~kOe/$\mu_{\rm B}$.\cite{Gavilano167202} Using the relevant parameters, $A_{\rm z} \simeq 1.1$\,kOe/$\mu_{\rm B}$, $\gamma_N = 103.962 \times 10^6\,{\rm rad}$~s$^{-1}$\,T$^{-1}$, $z=2$, $z^\prime=2$, $g=2.17$ [obtained from the $\chi (T)$ analysis in Ref.~\onlinecite{Ranjith2015}], $S=\frac12$, and the high-temperature (45\,K) relaxation rate of
$\left(\frac{1}{T_1}\right)_{T\rightarrow\infty}\simeq 113.37$\,s$^{-1}$ for the Li site, the magnitude of the maximum exchange coupling constant is calculated to be $J_{\rm max}/k_{\rm B}\simeq 13$\,K, which is in reasonable agreement with our leading exchange coupling ($J/k_{\rm B} = 15$~K) estimated from the band structure calculations.\cite{Ranjith2015}

As discussed before, $1/T_1$ decreases slowly for $T < 20$~K in contrast to the increase expected due to the growth of antiferromagnetic correlations. Similar kind of $T$-dependence has been observed in several antiferromagnets\cite{Nath214430,Bossoni014412,Carretta094420,Tsirlin014429} and the decrease in $1/T_1$ above $T_{\rm N}$ is explained by the partial filtering of antiferromagnetic fluctuations from the neighboring spins. In those compounds, the probed nucleus is coupled to nearest neighbor spins having opposite directions. In contrast, the Li nuclei in Li$_2$CuW$_2$O$_8$ are coupled to two nearest neighbor spins having the same direction. In this case, the filtering of antiferromagnetic fluctuations should not happen. Similar behaviour has been observed in frustrated breathing pyrochlore lattice compound LiInCr$_4$O$_8$, where the decrease in $1/T_1$ is explained by the opening of a spin gap above $T_{\rm N}$.\cite{Tanaka2014} In LiInCr$_4$O$_8$, $\chi(T)$ also shows an activated behaviour at low temperatures.\cite{Okamoto097203} However, in this compound $\chi(T)$ does not show any exponential decrease, thus ruling out the possibility of a spin gap at low temperatures. For spin-$1/2$ Heisenberg AFM spin chains, it has been predicted theoretically that the contribution of the staggered component ($q=\pm \pi/a$) to the spin-lattice relaxation rate behaves as $1/T_1 \sim T^{0}$ while the contribution of the uniform component ($q=0$) scales as $1/T_1 \sim T$.\cite{SandvikR9831,*Sachdev13006} The contributions of staggered and uniform components normally dominate in the low-temperature ($T\ll J$) and high-temperature ($T \sim J$) regimes, respectively.\cite{Nath174436} Our experimentally observed linear behavior of $1/T_1$ with temperature over the $5.5~K \leq T \leq 16$~K range in Fig.~\ref{T1}(a) is possibly due to 1D physics. Indeed, we found moderate 1D anisotropy of the 3D spin lattice in Li$_2$CuW$_2$O$_8$ with the leading interaction running along the $b$-direction.\cite{Ranjith2015} However, the purely 1D model is by far insufficient to describe the physics of the compound even in the $T\sim J$ temperature range.\cite{Ranjith2015}

At the 3D ordering temperature, the correlation length is expected to diverge, and $1/T_1$ in a narrow temperature range just above $T_{\rm N}$ (i.e., in the critical regime) should be described by the power law, $1/T_1 \propto \tau^{-\gamma}$, where $\gamma$ is the critical exponent. The value of $\gamma$ represents universal property of the spin system depending upon its dimensionality, the symmetry of the spin lattice, and the type of interactions. To analyze this critical behavior, $1/T_1$ is plotted against the reduced temperature $\tau$ in Fig.~\ref{T-Tn}. The data just above $T_{\rm N}$ ($\tau \leq 0.5$) were fitted by the power law with a fixed $T_{\rm N} \simeq 3.6$~K yielding $\gamma \simeq 0.10$.
For the 3D Heisenberg antiferromagnet, a mean-field theory predicts $\gamma = \frac12$ and a dynamic scaling theory gives $\gamma = \frac13$.\cite{Benner,Lee214416}
Our experimental value of $\gamma \sim 0.10$ is far below these theoretically predicted values. This effect requires further investigation and may be intertwined with the strong frustration in 3D that affects spin dynamics above $T_N$.

In the magnetically ordered state ($T<T_{\rm N}$), the strong temperature dependence of 1/$T_1$ is a clear signature of the relaxation due to scattering of magnons by the nuclear spins.\cite{Belesi184408} For $T\gg\Delta/k_{\rm B}$, 1/$T_1$ follows either a $T^3$ behavior or a $T^5$ behavior due to a two-magnon Raman process or a three-magnon process, respectively, where $\Delta$ is the energy gap in the acoustic magnon spectrum.\cite{Beeman359,Nath024431} For instance, in Pb$_2$VO(PO$_4$)$_2$, $1/T_1$ follows a $T^3$ behaviour\cite{Nath214430} while a $T^{5}$ behaviour has been reported for spin-$\frac12$ square-lattice compound Zn$_2$VO(PO$_4$)$_2$ (Ref.~\onlinecite{Yogi024413}) and the decorated Shastry-Sutherland lattice in the spin-$\frac12$ compound CdCu$_2$(BO$_3$)$_2$ (Ref.~\onlinecite{Lee214416}). On the other hand, for $T\ll\Delta/k_{\rm B}$, it follows an activated behavior $1/T_1 \propto T^2e^{-\Delta/k_{\rm B}T}$.

Clearly the $T^{3}$ behaviour couldn't describe our low temperature $1/T_1$ below $T_{\rm N}$ for Li$_{2}$CuW$_{2}$O$_{8}$. As one can see in Fig.~\ref{T1}(a), we fitted 1/$T_1$ below $T_{\rm N}$ by a $T^5$ behavior. An activated behavior would be consistent with our data too. However, the fit by $1/T_1 \propto T^2e^{-\Delta/k_{\rm B}T}$ in the temperature range 2~K to 3.3~K yields $\Delta \simeq 3.7$~K that is unusually large for this system and amounts to about 25\,\% of $J_{\max}$. 
This substantial spin gap is hard to justify for our system, because inversion symmetry of the crystal structure eliminates any Dzyaloshinsky-Moriya couplings, which are the leading anisotropy term in Cu$^{2+}$ oxides.\cite{[For example:][]rousochatzakis2015,*janson2014} The remaining anisotropy is of symmetric (Ising) type and does not exceed $1-2$\,\% in cuprates,\cite{lorenz2009} so it can not be responsible for the spin gap extracted from the fit of $1/T_1$. It is also worth noting that the specific heat follows the $T^3$ behavior at least down to 1.8~K (Fig.~\ref{heat-doped}). Therefore, we conclude that the $T^5$ behavior related to three-magnon processes is the most plausible scenario of the low-temperature spin-lattice relaxation, although measurements below 2~K could be useful to check whether the activated behavior might set in at very low temperatures corresponding to smaller values of the spin gap.


\subsection{Effect of doping}
\begin{figure}
\includegraphics {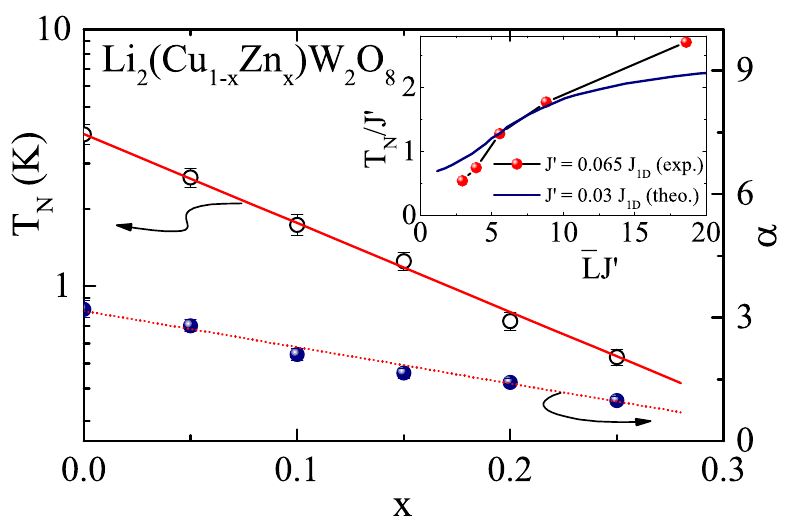}
\caption{\label{Tn} Log~$T_{\rm N}$ and $\alpha$ vs. $x$ obtained from the $C_{\rm p}(T)$ analysis in Fig.~\ref{heat-doped} in the left and right $y$-axes, respectively. The solid lines are the linear fits. Inset: $T_{\rm N}/J'$ vs. $\overline{L}J'$ where the average chain length $\overline{L} = 1/x-1$ and $J_{\rm 1D}=15$\,K is compared with the theoretical results from Ref.~\onlinecite{eggert2002}.}
\end{figure}
The spin lattice of Li$_2$CuW$_2$O$_8$ can be diluted by Zn atoms, and relatively high doping levels of $25-30$\,\% can be reached, which is not very common for Cu$^{2+}$ oxides because Zn$^{2+}$ does not favor the four-fold square-planar oxygen environment, which is triggered by the Jahn-Teller effect for Cu$^{2+}$. Other dopants may be compatible with this local environment of the Cu$^{2+}$ site, but their size difference prevents high doping levels. For example, only 3\,\% of Pd$^{2+}$ can be doped into the spin-chain antiferromagnet Sr$_2$CuO$_3$.\cite{kojima2004}

The linear evolution of $\theta_{\rm CW}$ upon doping (see the inset to Fig.~\ref{chi-doped}) confirms that Zn$^{2+}$ ions randomly break exchange bonds and reduce the overall exchange energy. This should reduce the ordering temperature $T_{\rm N}$, whereas the slope of $T_{\rm N}$ vs. $x$ depends on the nature of the system. Quantum systems, especially those having reduced dimensionality, will show higher slopes than their classical counterparts because the effect of dilution is amplified by quantum fluctuations that impede the long-range magnetic ordering. In spin-$\frac12$ square-lattice antiferromagnets, the slope of $T_{\rm N}(x)/T_{\rm N}(0)$ vs. $x$ is expected to be around 3.2.\cite{chernyshev2001} Even higher values were reported experimentally and ascribed to the combined effect of dilution, low-dimensionality and frustration induced by the doping.\cite{Carretta180411} Remarkably, none of those scenarios can be applied to Li$_2$CuW$_2$O$_8$, where the doping dependence of $T_{\rm N}$ can be tentatively linearized in the $\log T_{\rm N}$ vs. $x$ (and not in $T_{\rm N}$ vs. $x$) coordinates. This exponential reduction in $T_{\rm N}$ is strongly reminiscent of chain antiferromagnets. However, the model reported in the literature\cite{eggert2002} provides only a very crude fit of our data assuming a weak interchain coupling $J'/J_{\rm 1D}=0.065$ (see the inset of Fig.~\ref{Tn}). Comparing our data with the theory reported in Ref.~\onlinecite{zvyagin1975}, we conclude that Li$_2$CuW$_2$O$_8$ is similar to quasi-1D antiferromagnets with very weak interchain couplings, because the crossover between the exponential and linear decay of $T_{\rm N}$ is seen at $J'/J_{\rm 1D}\simeq 0.2$.

Naively, one would relate the 1D-like evolution of $T_{\rm N}$ upon doping to the presence of broad maxima in the magnetic susceptibility and specific heat. However, the weakly curved magnetization isotherm\cite{Ranjith2015} of Li$_2$CuW$_2$O$_8$ contradicts this 1D scenario and indicates strong interchain couplings that, according to our \textit{ab initio} results,\cite{Ranjith2015} reach $40-50$~\% of $J_{\rm 1D}$ rendering Li$_2$CuW$_2$O$_8$ a spatially anisotropic 3D antiferromagnet rather than a spin-chain antiferromagnet. Moreover, the ordered moment of 0.65~$\mu_B$ (Ref.~\onlinecite{Ranjith2015}) also indicates\cite{sandvik1999} interchain couplings of $J'/J_{\rm 1D}=0.4-0.5$ that are much stronger than $J'/J_{\rm 1D}=0.065$ expected from the doping dependence of $T_{\rm N}$.

Altogether, the doping experiments reveal peculiarities of Li$_2$CuW$_2$O$_8$. The evolution of $T_{\rm N}$ in doped samples gives insight into the nature of the parent compound, where strong magnetic frustration in 3D affects the ordered magnetic moment in the ground state\cite{Ranjith2015} and facilitates the suppression of $T_{\rm N}$ upon the dilution. The ground state of doped samples may be quite interesting on its own, and we note in passing that the power-law dependence of the low-temperature specific heat changes systematically from the conventional $T^3$ behavior in the parent compound to the nearly linear behavior at $x=0.25$. This effect warrants further investigation that, however, goes beyond the scope of our present study.

\section{Summary}
Frustrated nature of the 3D antiferromagnet Li$_2$CuW$_2$O$_8$ has been probed by $^7$Li NMR measurements and experiments on magnetically diluted Zn-doped samples. The double-horn spectral shape of the NMR line below $T_{\rm N}$ is attributed to the formation of commensurate magnetic order and the non-perpendicularity between the axis of the NMR coil and the resonance field. The absence of a narrow central line at the zero-shift position also supports earlier neutron diffraction results suggesting commensurate magnetic order with the reduced ordered moment. In contrast to the isostructural compounds Li$_2$NiW$_2$O$_8$ and Li$_2$CoW$_2$O$_8$, no magnetic transitions beyond that at $T_{\rm N}$ have been observed.

The nature of the 1/$T_1$ process below $T_{\rm N}$ remains somewhat ambiguous and a better analysis requires more data points in a wider temperature range below 2~K. The analysis of $1/\chi T_1 T$ suggests the frustrated nature of the compound. The critical exponent $\beta$ obtained from the sublattice magnetization below $T_{\rm N}$ is consistent with the 3D nature of the spin lattice. 

In Zn-doped sample, $T_{\rm N}$ decreases exponentially upon non-magnetic dilution $x$. This behavior is reminiscent of quasi-1D antiferromagnets and may be a footprint of magnetic frustration in the parent compound, because the quasi-1D scenario is clearly excluded by the size of the ordered moment and by the microscopic analysis. Heat capacity below $T_{\rm N}$ for all the doped samples follows the power law ($C_{\rm p} \propto T^{\alpha}$), but the exponent ($\alpha$) decreases from 3 to 1 as $x$ increases from 0 to 0.25. 

\acknowledgments
KMR and RN would like to acknowledge DST India for financial support. AAT was funded by the PUT733 grant of the Estonian Research Council, and by the Federal Ministry for Education and Research through the Sofja Kovalevskaya award of Alexander von Humboldt Foundation.

\end{document}